\documentclass[sigconf]{acmart}

\AtBeginDocument{%
  \providecommand\BibTeX{{%
    \normalfont B\kern-0.5em{\scshape i\kern-0.25em b}\kern-0.8em\TeX}}}

\usepackage[T5,T1]{fontenc}

\usepackage{sidecap}
\usepackage{wrapfig}

\newcommand{\BTP}{{\fontencoding{T5}\fontfamily{vietnam}\selectfont B\`ui T{\uhorn}{\`\ohorn}ng Phong}}
\newcommand{\Bui}{{\fontencoding{T5}\fontfamily{vietnam}\selectfont B\`ui}}
\newcommand{\Tuong}{{\fontencoding{T5}\fontfamily{vietnam}\selectfont T{\uhorn}{\`\ohorn}ng}}
\newcommand{\Nguyen}{{\fontencoding{T5}\fontfamily{vietnam}\selectfont Nguy{\~\ecircumflex}n}\;}
\newcommand{\Cam}{{\fontencoding{T5}\fontfamily{vietnam}\selectfont C\h\acircumflex m}}
\setcopyright{acmlicensed}
\copyrightyear{2018}
\acmYear{2018}
\acmDOI{XXXXXXX.XXXXXXX}

\acmJournal{TOG}
\acmVolume{37}
\acmNumber{4}
\acmArticle{111}
\acmMonth{8}

\setcopyright{rightsretained}
\copyrightyear{2024}
\acmYear{2024}
\acmConference{SIGGRAPH Talks '24}{July 27 - August 01, 2024}{Denver, CO, USA}\acmBooktitle{Special Interest Group on Computer Graphics and Interactive Techniques Conference Talks (SIGGRAPH Talks '24), July 27 - August 01, 2024}\acmDOI{10.1145/3641233.3664315}
\acmISBN{979-8-4007-0515-1/24/07}

\begin{document}

\title{The Life and Legacy of Bui Tuong Phong}


\author{Yoehan Oh}
\affiliation{%
  \institution{Yale University}  
  \country{USA}
  }
\email{yoehan.oh@yale.edu}

\author{Jacinda Tran}
\affiliation{%
  \institution{Harvard University}
  \country{USA}
}
\email{jtran@fas.harvard.edu}

\author{Theodore Kim}
\affiliation{%
  \institution{Yale University}
  \country{USA}
}
\email{theodore.kim@yale.edu}

\renewcommand{\shortauthors}{Oh, Tran, and Kim}

\begin{abstract}
We examine the life and legacy of pioneering Vietnamese computer scientist \BTP, whose shading and lighting models turned 50 last year. We trace the trajectory of his life through Vietnam, France, and the United States, and its intersections with global conflicts. Crucially, we present definitive evidence that his name has been cited incorrectly over the last five decades. His family name is {\Bui\;\Tuong}, not Phong. By presenting these facts at {\em SIGGRAPH}, we hope to collect more information about his life, and ensure that his name is remembered correctly in the future.
\end{abstract}

\maketitle

\section{Biography}


\begin{wrapfigure}{l}{0.325\columnwidth}
\centering
\vspace*{-1em}
\includegraphics[width=0.35\columnwidth]{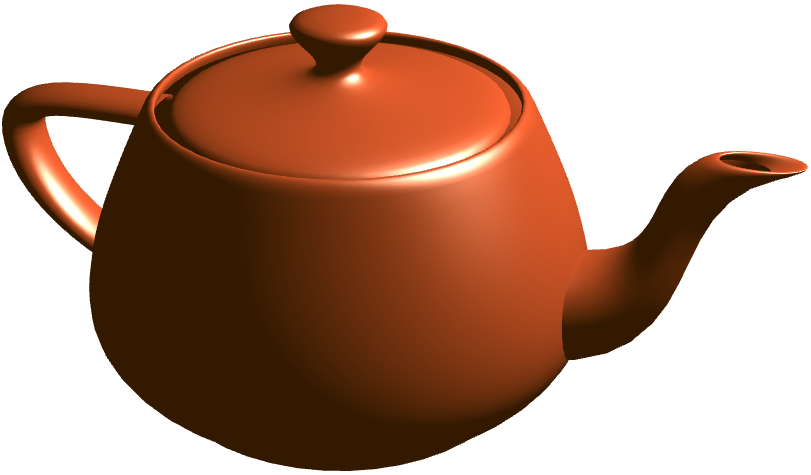}
\vspace*{-2em}
  \caption*{Phong shading and lighting on teapot.}
\vspace*{-1em}
\end{wrapfigure}
The Phong shading and lighting models \shortcite{phong1973illumination} turned 50 last year, and are ubiquitous in computer graphics. They are the most basic shaders in OpenGL and WebGL (see left), and run on browsers, video game consoles, and smartphones throughout the world. Despite this, very little is known about the life of their creator, \BTP.

\begin{wrapfigure}{r}{0.325\columnwidth}
\centering
\vspace*{-1em}
\includegraphics[width=0.35\columnwidth]{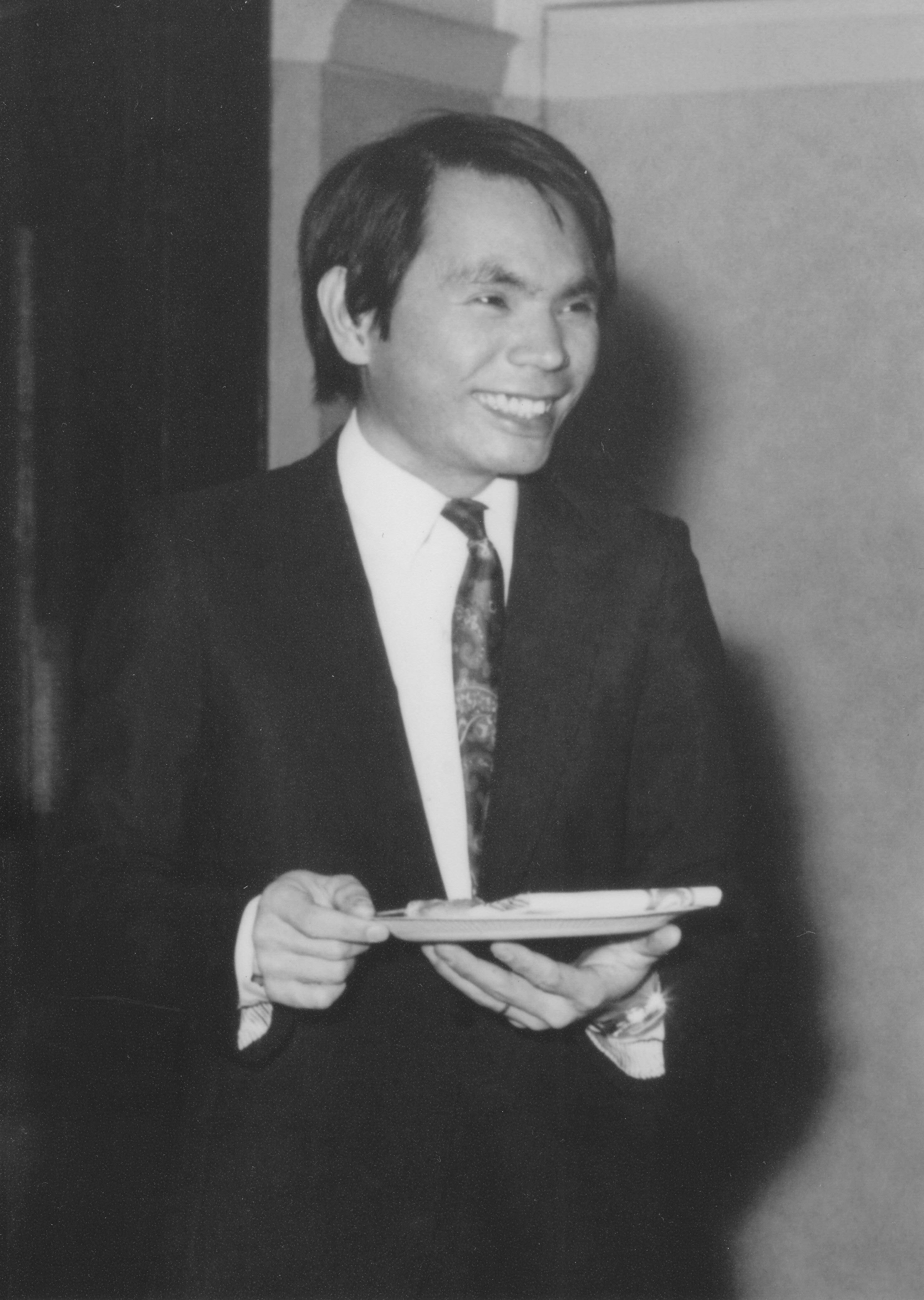}
\vspace*{-2em}
\caption*{Bui Tuong Phong, \textcopyright Dr.~Loan Hsieh}
\vspace*{-3em}
\end{wrapfigure}
As of July 2024, we are only able to locate one publicly available photo of his full face (see right), which we authenticated with his colleagues and family. Image searches for his name consistently and erroneously return images of the Vietnamese writer Chu \Cam\;Phong.


The most comprehensive information we have found is an article by computer scientist \Nguyen {\fontencoding{T5}\fontfamily{vietnam}\selectfont Ho\`ang}\;Thanh \shortcite{Nguyen2007}. In order to fill in the gaps and form a fuller picture of \BTP`s life, we will outline the facts discovered by \Nguyen here, then situate them within global events.

\Nguyen \shortcite{Nguyen2007} reports that \BTP\;was born in Hanoi in 1942. This was a turbulent time in Vietnam, as it was still a French colony (French Indochina), but the German invasion of France had left it vulnerable to Japanese occupation, and later on, indigenous uprising. \cite{kiernan2019}


\Nguyen reports that \BTP\;moved from Hanoi to Saigon in 1954. This coincides with the Geneva Conference, originally intended to broker the end of the Korean War, which partitioned the country into North and South Vietnam along the 17\textsuperscript{th} parallel. Over a million refugees fled from the Communist north (Hanoi) to the French south (Saigon) \cite{office2000state}.

\Nguyen reports that \BTP\;immigrated to France in 1964 and obtained several degrees before joining IRIA, present-day INRIA. The year of his immigration coincides with the Gulf of Tonkin incident, which dramatically escalated U.S. involvement in the Vietnam War, and initiated large-scale bombing of North Vietnam (\cite{kiernan2019} p.~432).

\Nguyen reports that \BTP\;immigrated to the U.S.~ in 1971, and obtained a Ph.D.~from the University of Utah in 1973. Various reports say he died of leukemia, lymphoma, or squamous cell carcinoma (SCC) in 1975, shortly after beginning a position at Stanford. In examining this final component, we must integrate some difficult facts from across his lifetime.

He lived in Saigon, present-day Ho Chi Minh City, from 1954 to 1964. This period overlapped with Operation Ranch Hand (1962-71), during which the U.S. military sprayed 20 million gallons of herbicide over Vietnam, Cambodia, and Laos. These rainbow herbicides, the most notable of which is Agent Orange, contain carcinogenic dioxins that are known to cause both leukemia and lymphoma.

\begin{SCfigure}[][b!]
\centering
\includegraphics[width=0.465\columnwidth]{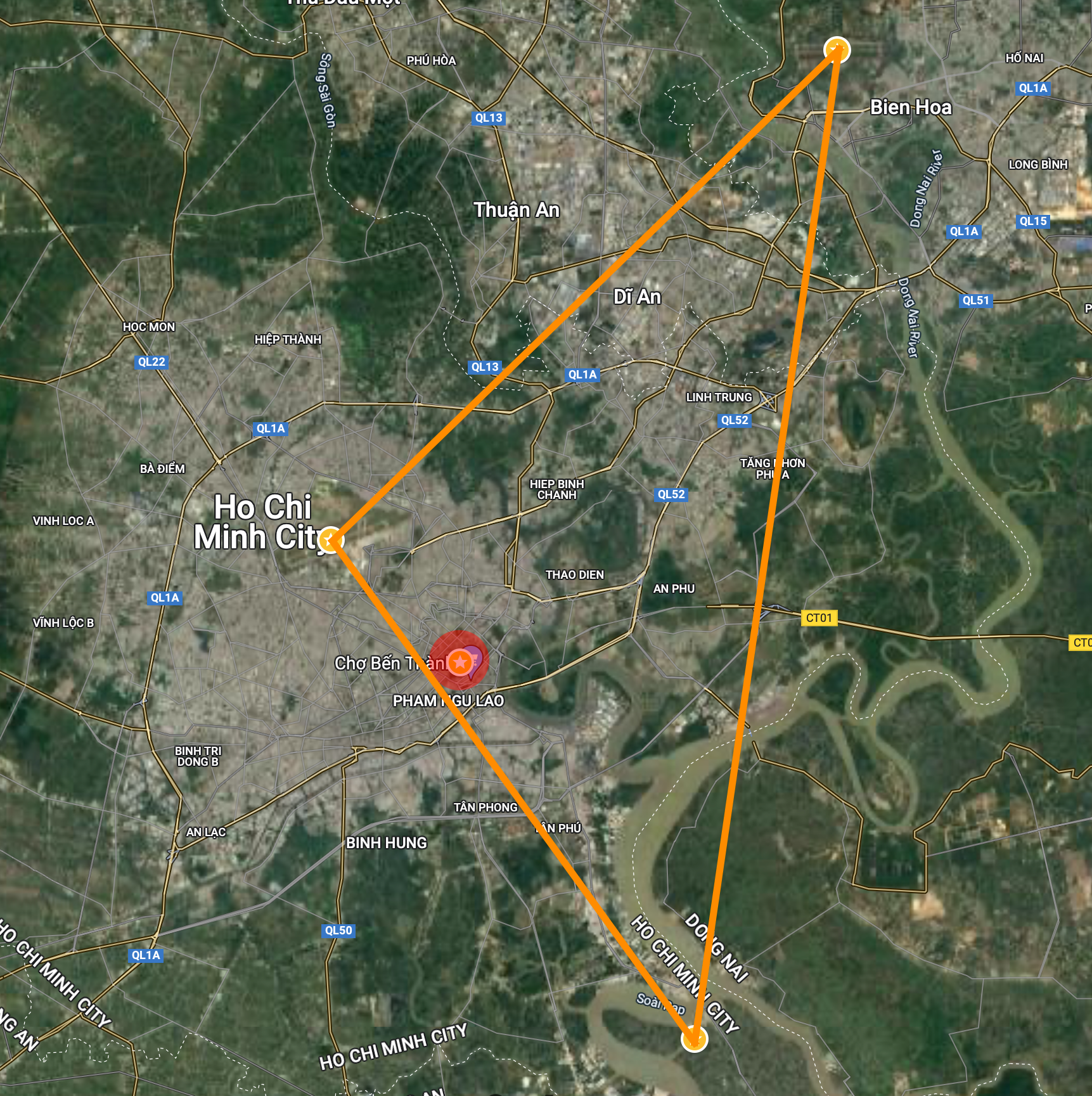}
\caption{Triangle of Ranch Hand sites, outside of Saigon in 1962. Clockwise from left: Tan Son Nhut and Bien Hoa air bases, and Than Tuy Ha ammunition dump (estimated from \citet{buckingham1982operation}). The red circle is {\em Lyc\'{e}e Jean Jacques Rousseau}, Dr.~Bui Tuong's school.}
\label{fig:triangle}
\end{SCfigure}

A variety of sites around Saigon are known to be spray sites or staging areas for Ranch Hand (Fig.~\ref{fig:triangle}). Two spray sites in 1962 were the U.S.~air base at Bien Hoa and the Than Tuy Ha ammunition dump (\citet{buckingham1982operation}, p.~35). The Agent Orange contamination at Bien Hoa is still being cleaned up today \cite{Stewart2018}. The Tan Son Nhut air base also served as a staging area and herbicide storage facility starting in 1962 (\citet{young2009history} p.~62).

Together, these sites form a triangle around the Lyc\'{e}e Jean Jacques Rousseau, the school that, according to \Nguyen, \BTP\;attended before moving to France in 1964. There is a high probability that he was exposed to the rainbow herbicides, though we will never know how much it contributed to his illness. For comparison, any U.S.~veteran who served in the Republic of Vietnam between 1962 and 1975 and later developed leukemia or lymphoma qualifies for \citet{VA2023} compensation. A dozen veterans have also successfully petitioned the VA for compensation for herbicide-related SCC.

While the tides of war flowed through \BTP's life, we are most interested in properly honoring his scientific achievements. Thus, we next turn to his legacy.

\section{Citation Confusion}

\BTP's death at an early age has created significant confusion around his name, particularly his given (first) name, and family (last) name. This sort of confusion was common with Vietnamese and other Asian names in the 1970s, because the Western convention is inverted, with the family name coming before the given name. The confusion appears to have carried over into the most basic component of his scientific legacy: his citations.

\subsection{The First Decade of Citations}

One of the earliest appearances of the term {\em Phong shading} is Ed Catmull's dissertation \shortcite{catmull1974subdivision}, where the bibliography lists {\em Bui Tuong-Phong} as its inventor. The hyphen suggests that {\em Tuong-Phong} is his ``compound'' given name \cite{Phan1985}, while {\em Bui} is his family name. The same hyphenation scheme is repeated in Fred Parke's dissertation \shortcite{parke1974parametric} that same year.

A year after his passing, the influential paper of \citet{blinn1976texture} shifted the hyphen to the first two syllables, {\em Bui-Tuong Phong.} That same year, \citet{max1976computer} abbreviated his name to {\em B.T. Phong,} further delineating {\em Phong} as the family name. 
The highly cited textbook by \citet{foley1982fundamentals} further inverted the name, and described the {\em shading model developed by Phong Bui-Tuong}\;(p.~577). The authors listed {\em Bui-Tuong} as his family name, and the citation became {\em [BUIT75] Bui-Tuong, Phong}\;(p.~628). 

In the first decade after \BTP's passing, citations variously listed his family name as {\em Bui}, {\em Phong}, and {\em Bui-Tuong}. His CACM article \shortcite{Bui1975}, cited over 5000 times as of this writing, offers no additional guidance. He listed his own name without any hyphens: {\em Bui Tuong Phong}. Today, in the ACM Digital Library, his citations appear under the name {\em Phong, Bui Tuong}.

\subsection{{\em Bui Tuong} Appears to be Correct}

\BTP's family name is highly unlikely to be {\em Phong}, because {\em \Bui} is an extremely common family name in Vietnam, e.g.~like {\em Smith} or {\em Miller} in the U.S. To draw a Western analogy, while it is certainly possible that {\em John Smith} refers to a person named {\em Smith} whose last name is {\em John}, it is highly unlikely.

After we published a version of our findings in a popular article \cite{Kim2024} speculating that his family name was {\em \Bui}, we were contacted by Dr.~Loan Hsieh (ne\'{e} Bui Tuong), \BTP's daughter. She informed us that the name \Bui\;is so common that \BTP's father added his middle name (\Tuong) to the last name (\Bui) ``to make it stand out.'' In an email on May 31, 2024, she wrote: ``Our family name is Bui Tuong (sometimes with or without the hyphen which makes growing up in the United States an identity crisis for all of us). My father’s first name is Phong.''




The citation should be: {\em Bui Tuong, Phong}. The Western ordering, in the style of {\em John Smith}, should be: {\em Phong Bui Tuong}.


\subsection{{\em Phong Shading} Appears to be Correct}

The preceding also implies Dr.~\Bui\;\Tuong's shading model has been named incorrectly; i.e.~it should be the {\em Bui Tuong} model, not the {\em Phong} model. However, if we continue to adopt the naming conventions that Dr.~\Bui\;\Tuong\;himself used, this does not appear to be the case.

In Figure 4.12 of his dissertation \shortcite{phong1973illumination}, he directly refers to his shading model as {\em Phong Improved Shading.}  The abstract of his final publication \cite{phong1975improved} additionally refers to {\em Phong and Gouraud shading.}

We have not been able to determine whether he himself started calling it the {\em Phong} model, or whether the eponym originated from one of his colleagues, and he then went along with it. Given how eponyms are usually coined, the latter seems more likely. Our hope is that a {\em SIGGRAPH} attendee might have first-hand knowledge of this, because our exchanges with early researchers (Gouraud, Catmull, Clark, Sutherland) have not yielded anything definitive.

Regardless, Dr.~\Bui\;\Tuong\;did not contest the naming. His personal reasons will likely remain unknown, but one possibility is that he saw its poetry: {\em phong} translates to {\em (of the) wind}.


\bibliographystyle{ACM-Reference-Format}
\bibliography{sample-base}

\end{document}